\documentclass{nature}
\usepackage[squaren]{SIunits}
\usepackage{lineno}
\usepackage{graphicx}

\title{Femtosecond probing of light-speed plasma wakefields by using a relativistic electron bunch}

\author{C.J. Zhang$^{1,2,7}$, J.F. Hua$^1$, Y. Wan$^1$, B. Guo$^1$, Y.P. Wu$^1$, C.-H. Pai$^1$, F. Li$^1$, H.-H. Chu$^3$, Y.Q. Gu$^2$, X.L. Xu$^4$, W.B. Mori$^4$, C. Joshi$^4$, J. Wang$^{3,5,6,*}$, W. Lu$^{1,7,*}$}

\begin{document}
\maketitle

\begin{affiliations}
 \item Department of Engineering Physics, Tsinghua University, Beijing 100084, China
 \item Laser Fusion Research Center, China Academy of Engineering Physics, Mianyang, Sichuan 621900, China
 \item Department of Physics, National Central University, Jhong-Li 32001, Taiwan
 \item University of California Los Angeles, Los Angeles, California 90095, USA
 \item Institute of Atomic and Molecular Sciences, Academia Sinica, Taipei 10617, Taiwan
 \item Department of Physics, National Taiwan University, Taipei 10617, Taiwan
 \item IFSA Collaborative Center, Shanghai Jiao Tong University, Shanghai 200240, China
\end{affiliations}

\begin{abstract}
Relativistic wakes produced by intense laser or particle beams propagating through plasmas are being considered as accelerators\cite{Tajima_Laser_1979,Chen_Acceleration_1985} for next generation of colliders and coherent light sources\cite{Esarey_Physics_2009}. Such wakes have been shown to accelerate electrons and positrons to several gigaelectronvolts (GeV)\cite{Hogan_Multi_2005,Blumenfeld_Energy_2007,Kim_Enhancement_2013,Litos_High_2014,Corde_Multi_2015,Wang_Quasi_2013,Leemans_Multi_2014}, with a few percent energy spread\cite{Corde_Multi_2015,Wang_Quasi_2013,Leemans_Multi_2014} and a high wake-to-beam energy transfer efficiency\cite{Litos_High_2014}. However, complete mapping of electric field structure of the wakes has proven elusive. Here we show that a high-energy electron bunch can be used to probe the fields of such light-speed wakes with femtosecond resolution. The highly transient, microscopic wakefield is reconstructed from the density modulated ultra-short probe bunch after it has traversed the wake. This technique enables visualization of linear wakefields in low-density plasmas that can accelerate electrons and positrons beams. It also allows characterization of wakes in plasma density ramps critical for maintaining the beam emittance\cite{Xu_Physics_2016,Mehrling_Transverse_2012}, improving the energy transfer efficiency\cite{Litos_High_2014,Lu_Generating_2007} and producing high brightness beams\cite{Geddes_Plasma_2008,Plateau_Low_2012} from plasma accelerators.
\end{abstract}

When an intense laser or charged particle beam passes through a plasma it leaves behind a wake caused by the charge separation of plasma electrons and ions\cite{Tajima_Laser_1979,Chen_Acceleration_1985}. The phase velocity of the wake is nearly the speed of light \emph{c} and the longitudinal electric field can be orders of magnitude larger than that in conventional accelerators. To optimize both the efficiency and the energy spread of the accelerated beam, accurate information about the field structure of the wake is needed. This has proved difficult because the wakes are microscopic, highly transient and relativistic. In previous studies, optical methods such as frequency domain interferometry\cite{Matlis_Snapshots_2006} and ultra-fast shadowgraphy\cite{Buck_Real_2011,Sävert_Direct_2015} were developed to diagnose the dimensions and shape of the wake in high density plasmas. Recently the transverse oscillations of the drive electron beam itself have been shown to give spatially integrated information about the longitudinal variation of the fields of a highly nonlinear wake\cite{Clayton_Self_2016}. No technique has hitherto successfully given instantaneous information about both the transverse and longitudinal distribution of the electric field of the wake. Linear (sinusoidal) plasma wakes are being explored for accelerating both electrons and positrons- \emph{a necessary condition for an electron-positron collider}\cite{Esarey_Physics_2009}. Here we show that a femtosecond (fs) duration high-energy bunch of electrons generated by a laser wakefield accelerator (LWFA) can be used to take a snapshot of a second relativistically propagating laser-driven linear wake to give information about its instantaneous longitudinal and transverse field structure as well as its spatio-temporal evolution.

\begin{figure}
\includegraphics[width=5.5in]{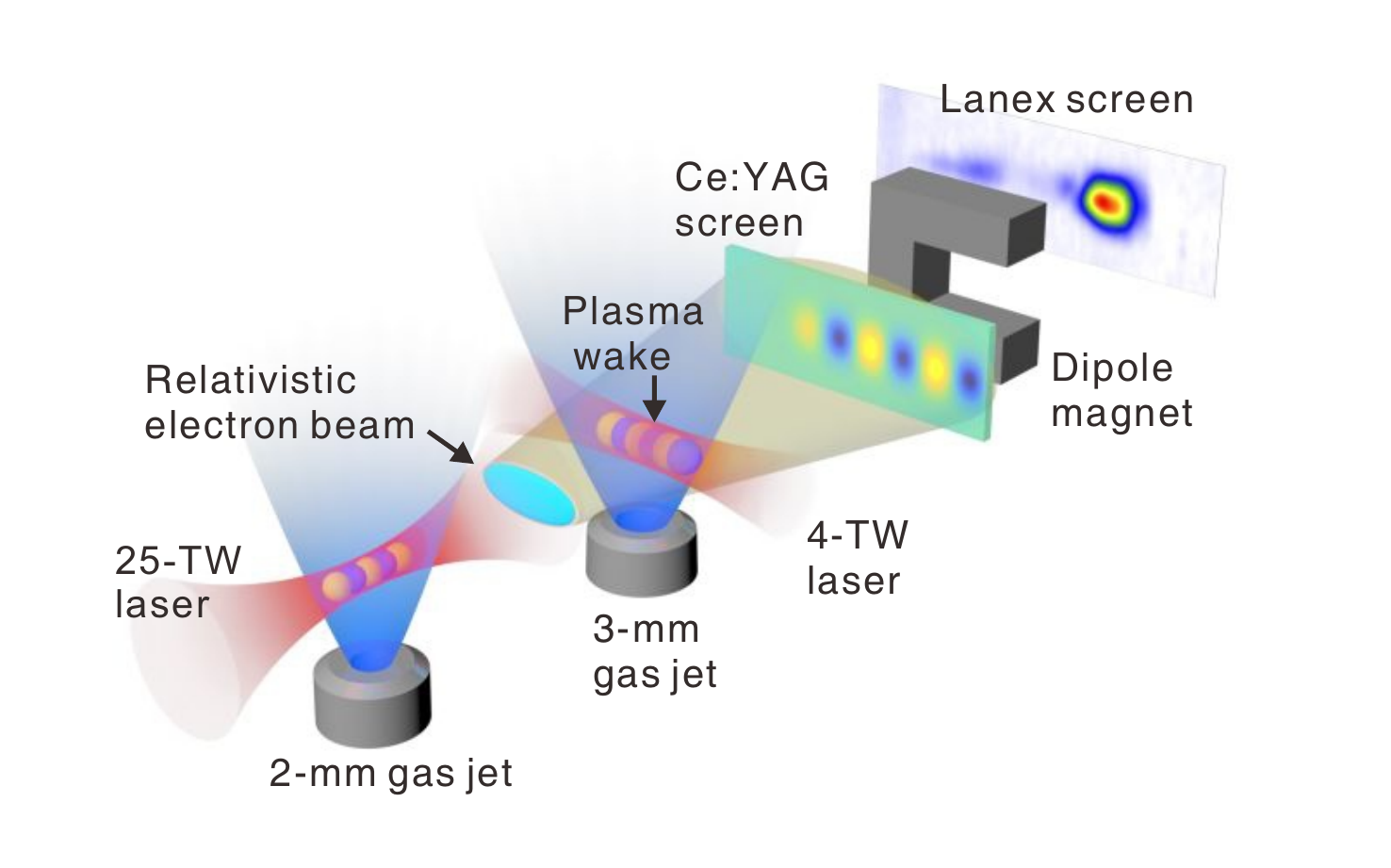} 
\caption{The schematic drawing of the probing of a laser-driven plasma wakefield using a relativistic electron beam.}
\label{fig1}
\end{figure}

The concept is illustrated in Fig. 1 and is described fully in the supplementary information (see Supplementary Information Section 1, SI1). Briefly, the relativistic probe beam is generated by focusing a 25-$\tera\watt$, $\sim$40-$\femto\second$ (full width at half maximum, FWHM), Ti-sapphire laser pulse to give a normalized vector potential $a_0$ of $\sim$2 into a $8\times10^{18}~\centi\meter^{-3}$ density plasma produced using a gas mixture of 95\% helium and 5\% nitrogen. The laser pulse excites a highly nonlinear wake and accelerates electrons via ionization trapping of nitrogen K-shell electrons\cite{Pak_Injection_2010} to give energies between 60-80 $\mega\electronvolt$ with a 15-$\mega\electronvolt$ energy spread and a charge of 2-10 $\pico\coulomb$. The pulse length of the probe is $\sim$4 $\femto\second$ (FWHM)\cite{Zhang_Temporal_2016}, the divergence is $\sim$7 $\milli\radian$ (FWHM) (see SI2). After propagating through free space for 11 $\centi\meter$ the probe beam intercepts at right angle a second weaker/linear wake produced by a 4-$\tera\watt$, 100-$\femto\second$, $a_0\approx0.25$ Ti-sapphire laser in a variable density helium plasma. Both plasmas are produced by ionization of the gas emanating from supersonic gas jet nozzles. As the electron probe traverses the second plasma, some of the electrons are deflected by the electric field of the wake. These deflections, which originally appear as transverse momentum modulations of the probe beam, evolve into density modulations. The density modulated probe is detected by recording the radiation produced by the beam impinging upon a 100 $\micro\meter$ thick cerium-doped yttrium aluminium garnet (Ce:YAG) screen (see Methods). This screen can be removed online to allow the probe beam to be dispersed by a magnetic spectrometer to record the energy spectrum of the electrons.

\begin{figure}
\includegraphics[width=5.5in]{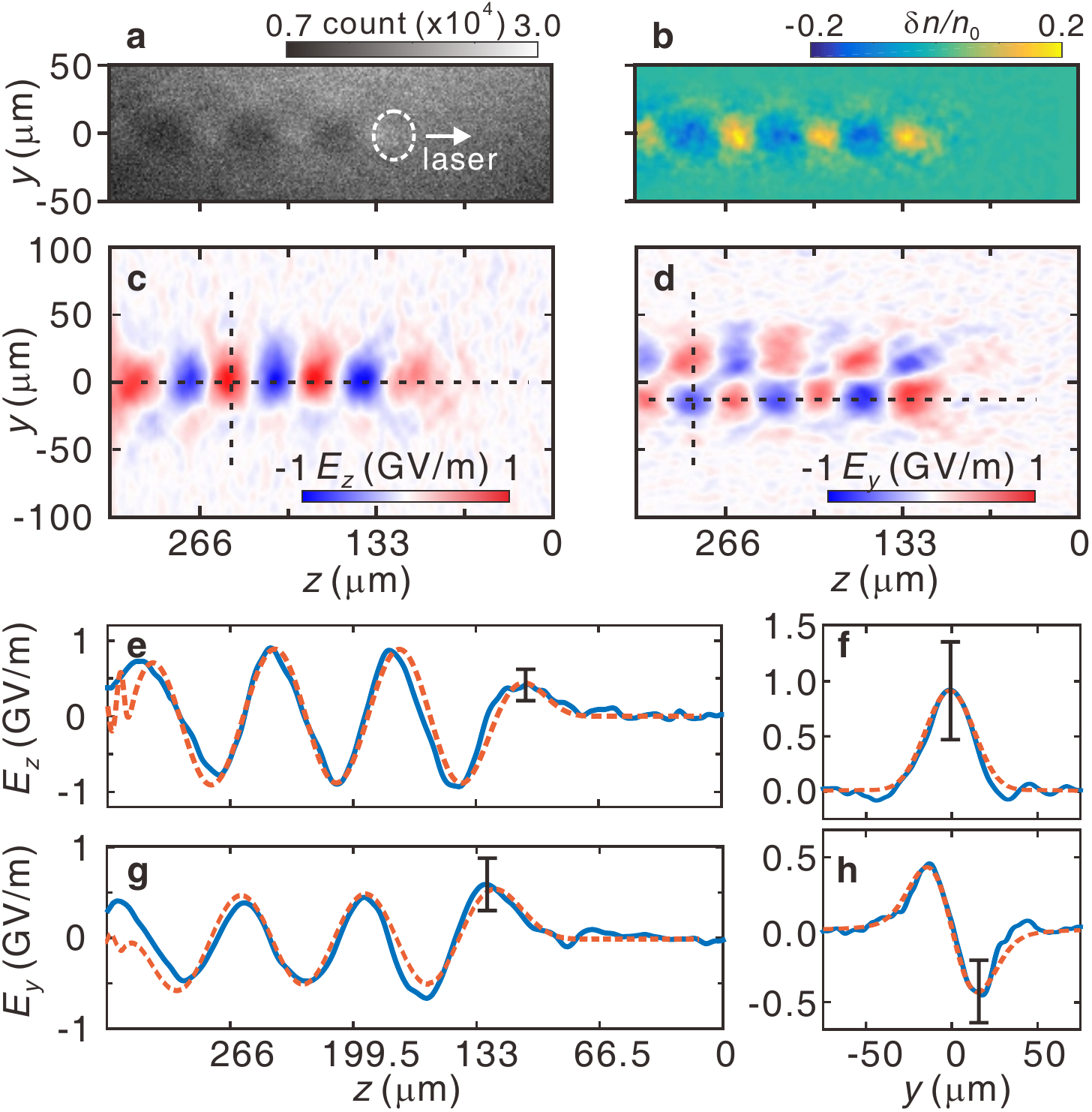} 
\caption{Snapshot of the wake, the deduced electric field structure of the wake and comparison with simulations. \textbf{a}, Raw image of the probe density $n$. \textbf{b}, Density modulation $n/n_0$. \textbf{c} and \textbf{d}, The $(y,z)$ plot of the reconstructed $E_z$ and $E_y$ fields respectively. \textbf{e} and \textbf{f}, The axial and transverse ($z=242~\micro\meter$) lineouts of $E_z$ respectively. \textbf{g} and \textbf{h}, The axial (at $y=-10~\micro\meter$) and transverse ($z=300~\micro\meter$) lineouts of $E_y$. The blue solid lines show the experimental data, while the red dashed lines are from simulations. The error bars arise from the fluctuating energy, pulse width and correlated divergence of the probe beam.}
\label{fig2}
\end{figure}

By properly adjusting the timing between the electron probe and the 4-$\tera\watt$ laser, a snapshot of the wake is obtained on the Ce:YAG screen as shown in Fig. 2a. Figure 2b shows the density modulation $\delta n/n_0=(n-n_0)/n_0$ of the probe derived from this snapshot (see Methods and SI3). The wavelength of the wake being probed is $\sim$$65~\micro\meter$, in very good agreement with the wavelength of 61 $\micro\meter$ deduced from the plasma density of $3\times10^{17}~\centi\meter^{-3}$ measured by interferometry (by extrapolating the data measured at higher backing pressures). The radius of the wake is about 10 $\micro\meter$ (root-mean-square, rms) or $\sim$one collisionless skin depth of the plasma. The transit time of the probe through the wake is 66 $\femto\second$. However, the time resolution is mainly determined by the length of the probe beam (4 $\femto\second$) for a linear wake as shown in ref. 22.

A theoretical model\cite{Zhang_Capturing_2016} is developed to relate $\delta n/n_0$ with the electric field $E(x,y,z)$ of the wake,
\begin{equation}
\label{eqn:1}
\frac{\delta n}{n_0}=\frac{K_\tau K_E K_\theta}{M}\frac{eL}{\beta cp_0}\nabla\cdot\int_{-s}^s E(x,y,z-\beta_Ex)dx
\end{equation}
where the integral limits $\pm s$ should be large enough to cover the transverse extent of the wakefield\cite{Zhang_Capturing_2016}. Here the wake propagates along the \emph{z} direction at a speed of $\beta_Ec$ and the probe traverses along the \emph{x} direction: $p_0=\gamma m_ec$ is the central momentum ($\gamma\approx140$) of the probe beam, \emph{e} is the electron charge, $\beta c$ is the velocity of the probe, $L=42~\centi\meter$ is the drift distance in vacuum, $M=4.8$ is the geometric magnification, $K_\tau$, $K_E$ and $K_\theta$ are correction factors for pulse length, energy spread and correlated divergence angle of the probe beam.

With the measured density modulation in hand, we can then reconstruct the wakefield by solving equation (1) (see Methods). The reconstructed fields for this shot are shown in Figs. 2c and d, for the longitudinal $E_z$ and transverse $E_y$ component of the electric field respectively from the density modulation. 

To make a quantitative comparison, we take axial and radial lineouts of the longitudinal and transverse fields as shown in Figs. 2e through h. Since the laser is weak ($a_0\approx0.25$) and has a near Gaussian profile, the wake being probed is quasi-linear and therefore the longitudinal field is expected to have a near-sinusoidal form along the \emph{z} direction and a Gaussian form along the \emph{y} direction\cite{Esarey_Physics_2009} as seen in the blue experimental curves in Figs. 2e and f. Furthermore $E_z(y,z)$ obtained from the particle-in-cell (PIC) simulation\cite{Fonseca_OSIRIS_2002} (red dashed lines) (see Methods) fit both the observed longitudinal and transverse variations of the measured $E_z$ very well. Note that the simulation used an $a_0=0.22$ to account for the non-ideal laser focal spot in the experiment. Figures 2g and f show the reconstructed longitudinal and transverse variations of the $E_y$ field. In the longitudinal direction, the $E_y$ field is $\pi/2$ out of phase with the $E_z$ field as expected for a linear wake\cite{Esarey_Physics_2009}. Meanwhile, the transverse variation of the $E_y$ field is of the form $ye^{-y^2/2\sigma^2}$\cite{Esarey_Physics_2009}, also in excellent agreement with the measured $E_y$ field. The peak accelerating field is $\sim$$0.9\pm0.4~\giga\electronvolt\per\meter$, in good agreement with simulations. Moreover, the transverse field is approximately linear over $\pm8~\micro\meter$. This linear portion is critical for preserving the emittance of the accelerating electrons. Although the wake measured in this experiment is in the linear regime, this probing technique can be extended to give qualitative features of highly nonlinear wakes\cite{Zhang_Capturing_2016}.

We now demonstrate the application of this technique to characterizing plasma wakes in density ramps. To maximize the charge throughput and minimize the emittance growth, the accelerating bunch must be carefully matched from one plasma accelerator stage to the next. This can be done in principle by employing plasma density ramps at the entrance to and the exit from the plasma acceleration section to properly increase/decrease the focusing force acting on the electrons\cite{Xu_Physics_2016,Mehrling_Transverse_2012}. Wakes produced in plasma density up-ramps may be useful for increasing the dephasing (or acceleration) length to match the pump depletion length of the laser pulse\cite{Lu_Generating_2007} thereby increasing the energy transfer efficiency of a LWFA. On the other hand, density downramps have been suggested for reducing the phase velocity to controllably inject charge into the wake to generate high brightness beams\cite{Geddes_Plasma_2008,Plateau_Low_2012}. However, detailed mapping of the wakes in such low-density plasma ramps has not been done with existing techniques.

\begin{figure}
\includegraphics[width=5.5in]{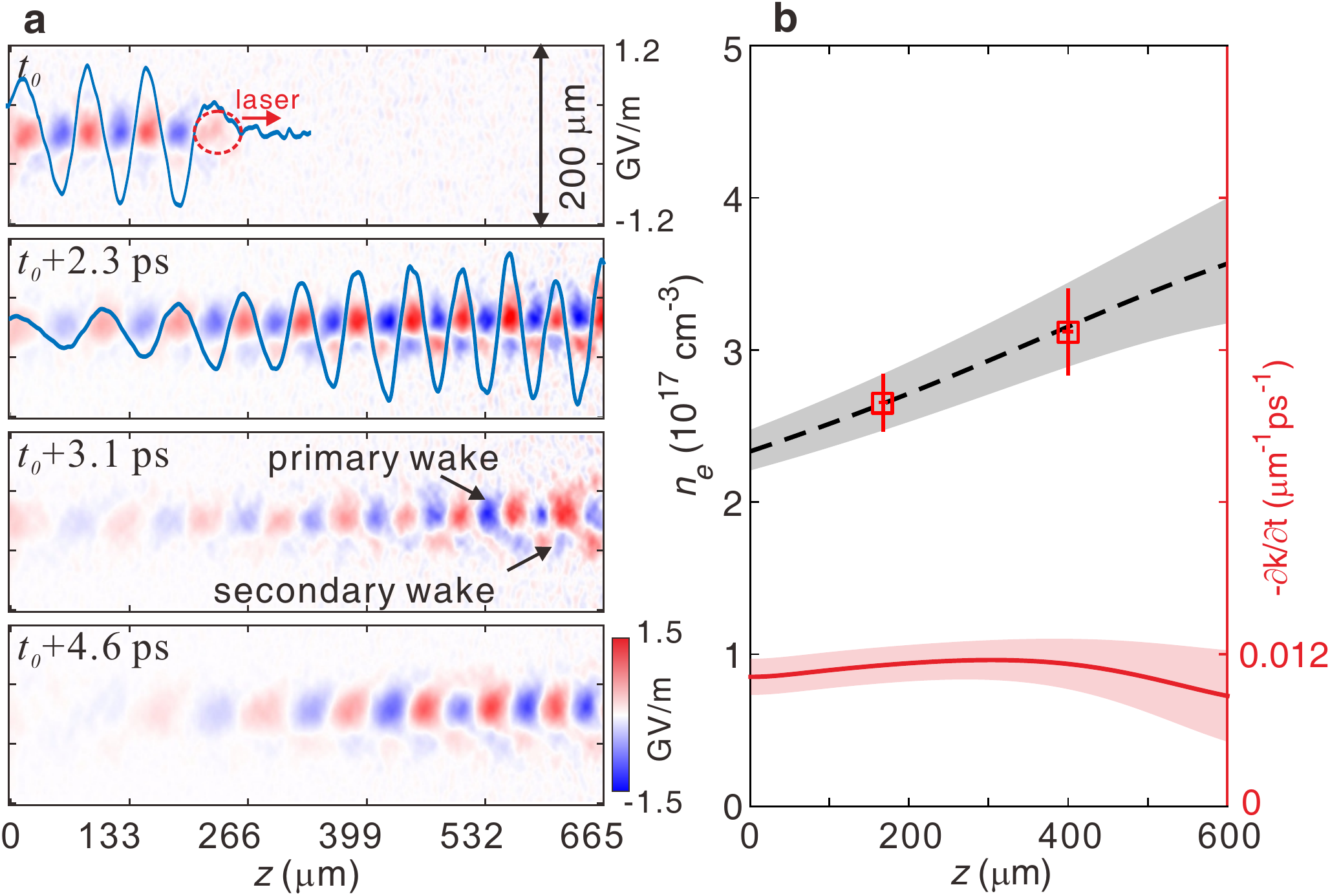} 
\caption{Spatio-temporal evolution of the wakefield in an increasing plasma density ramp. \textbf{a}, Four snapshots showing the evolution of the $E_z(z, t)$ field. The blue curves show the wavelength variation of the wake. During this run the laser spot had a double lobed structure that produced a weaker secondary wake about $\pi$ radians out of phase with the primary wake as seen in the third frame in Fig. 3a (see SI4). \textbf{b}, The deduced density profile of the plasma as seen by the laser pulse (black dashed line). The two red points are direct estimation of the density from $\omega_p=ck_p$ using the value of $k_p(t=0, z_1=170~\micro\meter, z_2=400~\micro\meter)$ just behind the laser pulse while the laser pulse was still visible in the frame (see SI5). The vertical spread is from shot-to-shot density variation and the horizontal errorbar is from the uncertainty in determining the location of the laser pulse. The red curve in \textbf{b} is the value of $-\partial k/\partial t=\partial\omega_p/\partial z$ from the data in \textbf{a} using which the density profile shown above is obtained. The shaded regions represent the uncertainty in measuring $-\partial k/\partial t$ at different $z$ values.}
\label{fig3}
\end{figure}

By varying the time delay between the two laser pulses and by focusing the weaker laser at the end of the rising density ramp that precedes uniform density region of the gas jet, the spatio-temporal evolution of the wake (28 shots at 8 different times) was recorded, as shown in Fig. 3a. The wake wavelength $\lambda$ decreases from the left to the right in Fig. 3a in all the frames due to the density gradient. Furthermore, at a particular \emph{z} location $\lambda$ increases with time. These two observations are not attributable to plasma expansion\cite{Joshi_Resonant_1982} which occurs over a longer timescale but instead can be explained as follows. First we note that a cold plasma wake has zero group velocity and continues to oscillate at its local plasma frequency $\omega_p(z)$ after the laser pulse has passed. The phase of the wave is therefore $\phi=\omega_p(z)(t-z/v_d)$ where $v_d\approx c$ is the velocity of the drive pulse. Therefore, $\omega(z,t)\equiv\frac{\partial\phi}{\partial t}=\omega_p(z)$ and $k(z,t)\equiv-\frac{\partial\phi}{\partial z}=k_p(z)-\frac{\partial\omega_p}{\partial z}t$ from which it follows that the wavenumber evolves in time according to $-\partial k/\partial t=\partial\omega_p/\partial z$\cite{Whiteman_Linear_1975}. This change in \emph{k} happens at every \emph{z} location as time passes. Therefore by measuring $k(z)$ for 8 shots that are roughly $0.77~\pico\second$ apart, we obtain $-\partial k(z)/\partial t$ (red curve in Fig. 3b) and use this to obtain $\omega_p(z)\equiv-\int\frac{\partial k(z)}{\partial t}dz$ (see SI5). This can be converted into $n_e(z)=m_e\varepsilon_0\omega_p^2/e^2$ by noting that in the top frame in Fig. 3a the laser pulse has just reached the point $z\approx170~\micro\meter$, therefore $k=k_p(t_0)=\omega_p/c$. This is shown in Fig. 3b by the black curve where the plasma density is seen to rise from $\sim$$2.3\times10^{17}~\centi\meter^{-3}$ to $\sim$$3.6\times10^{17}~\centi\meter^{-3}$ in $600~\micro\meter$. This temporal variation of the wavenumber of the wake has been confirmed in PIC simulations with mobile ions.

In conclusion, we have demonstrated the imaging of light-speed plasma wake with femtosecond resolution and the reconstruction of its field structure by utilizing an ultra-short relativistic electron probe. Such complete information obtained in the experiments is vital to the development of plasma based wakefield accelerators.

\begin{methods}
\subsection{Imaging system.}
The density modulation of the electrons after 42-$\centi\meter$ of free space drift is converted into visible light signals by a 100-$\micro\meter$ thick Ce:YAG screen. The visible light is collected and transmitted by the imaging system consisting of a coated silver mirror and a plano-convex lens pair and then recorded by an electron multiplying charge coupled device (EMCCD). The first plano-convex lens has a focal length of $f=180~\milli\meter$ while the second lens has a focal length of $f=600~\milli\meter$. The magnification factor of the imaging system is 2.5. The imaging system is placed on a two-dimensional translation stage to ensure high spatial resolution, which is measured to be $2.8~\micro\meter$ (1.16 times diffraction limit). The imaging system can be removed online so that the spectrum of the electron bunch can be measured using the electron energy spectrometer.
\subsection{Field structure reconstruction.}
To reconstruct the electric field structures, we need to know the density modulation $\delta n/n_0$ where $\delta n=n-n_0$, and $n$ is the measured probe density, $n_0$ is the corresponding background density of the probe beam. The background density is obtained from the corresponding data itself by filtering out the high frequency components in the data. The background is then subtracted from the raw data to give $\delta n=n-n_0$ and the result is divided by $n_0$ to give $\delta n/n_0$. Ref. 22 gives a detailed procedure for retrieving the electric field $E(x,y,z)$ from the corrected density modulation ($\delta n/n_0$ divided by the correction factors $K_\tau$, $K_E$ and $K_\theta$). For the plasma and probe beam parameters used in the experiment, $K_\tau\sim1$, $K_E\sim1$ and $K_\theta\sim0.3$ (see SI3). Here we summarize the key steps: Step 1. Decoupling the $E_z$ and $E_y$ components of the electric field of the wake in equation (1) by using the Panofsky-Wenzel theorem\cite{Panofsky_Some_1956} which states that $\partial E_z/\partial y=\partial E_y/\partial z$ to get two separate equations for $E_z$ and $E_y$ respectively (similar to Eq. 7 and Eq. 8 in ref. 22). These two equations relate the partial derivative of the measured density modulation $I\equiv\delta n/n_0$ along \emph{z} (the wake propagation direction) and \emph{y} (the orthogonal direction) direction, i.e., $\partial I/\partial z$ and $\partial I/\partial y$ to the longitudinal field $E_z$ and transverse field $E_y$ respectively. Step 2. Solving these two Poisson's equations to get the integral terms, which are Abel transforms of the integrands. Step 3. Instead of performing the standard Abel inversion, we multiply a factor of $K=Mp_0L^{-1}(\sqrt{2\pi}\sigma_E)^{-1}\exp((k\sigma_E)^2/2)$ to the results obtained in the previous step to get the final $E_z$ and $E_y$ to suppress the noise introduced in Abel inversion. This is based on the fact that the Abel transform of a Gaussian function $\exp(-r^2/2\sigma_E^2)$ preserves its form thus can be simplified by multiplying a factor of $\sqrt{2\pi}\sigma_E$ to the original function. The factors in $K$ are defined as follows: $M=4.8$ is the geometric magnification factor, $p_0$ is the initial momentum of the probe, $L$ is the free drift distance, $k$ is the local plasma wave number and $\sigma_E$ is the local radius of the wake. Both the local wake wavenumber $k$ and wake radius $\sigma_E$ can be directly obtained from the measured density.
\subsection{Particle-in-cell simulation.}
The simulation results shown in Fig. 2 are obtained with the code OSIRIS\cite{Fonseca_OSIRIS_2002}. The simulation box is $300~\micro\meter$ along the laser propagation direction (\emph{z}) and $200~\micro\meter$ along the orthogonal direction (\emph{y}). The box is divided into 11800 and 1570 cells along the respective directions. Four particles are initialized in each cell. Pre-ionized plasma is used. The profile of the plasma starts with a $12.7~\micro\meter$ linearly rising edge and followed by a plateau. The peak density of the plasma is $3\times10^{17}~\centi\meter^{-3}$. A 100-$\femto\second$ (FWHM), p-polarized laser pulse with a sin$^2$ intensity profile is emitted from the left wall of the simulation box and is focused into the plasma with a spot size of $26.5~\micro\meter$ to give a vacuum $a_0$ of 0.22. The focal plane of the laser is placed at the right boundary of the simulation box. Absorbing boundary conditions are used for the electro-magnetic fields and particles in all directions. 
\end{methods}


\bibliographystyle{naturemag}


\begin{addendum}
 \item The authors thank Dr. Te-Sheng Hung, Mr. Ying-Li Chang, Mr. Yau-Hsin Hsieh and Mr. Chen-Kang Huang for helping this experiment. This work was supported by the National Basic Research Program of China No. 2013CBA01501, NFSC Grant No. 11425521, No. 11535006, No. 11175102, No. 11005063, No. 11375006 and No. 11475101, the Foundation of CAEP No. 2014A0102003, Tsinghua University Initiative Scientific Research Program, the Thousand Young Talents Program, the U.S. DOE Grant No. DE-SC0010064, No. DE-SC0008491, No. DE-SC0008316, No. DE-SC0014260, the U.S. NSF Grant PHY-1415386, No. ACI-1339893 and No. PHY-500630, and the Ministry of Science and Technology of Taiwan under Contracts No. 104-2112-M-001-030-MY3. Simulations are performed on Hoffman cluster at UCLA and Hopper, Edison cluster at National Energy Research Scientific Computing Centre (NERSC).
 \item[Author Contributions] 
 W.L. and J.W. conceived the experiment and provided the support and guidance of the project. C.J.Z designed the experiments. C.J.Z., J.F.H., Y.W., B.G., Y.P.W., C.H.P. and F.L. carried out the experiments. H.H.C. provided support on operating the laser system. C.J.Z, C.J., W.B.M. and X.L.X. did the simulation and data analysis. C.J.Z, J.F.H., C.H.P., C.J., J.W., W.L. W.B.M. and Y.Q.G. discussed the results and contributed to the manuscript.
 \item[Competing Interests] The authors declare that they have no competing financial interests.
 \item[Correspondence] Correspondence and requests for materials
should be addressed to \\ W.L. (e-mail: weilu@tsinghua.edu.cn) and J.W. (e-mail:jwang@ltl.iams.sinica.edu.tw).
\end{addendum}

\end{document}